# Deep Learning-based Power Allocation in Rate Splitting Optical Wireless Networks

Khulood D. Alazwary[1], Ahmad Adnan Qidan[2], T. E. H. El-Gorashi[2] and Jaafar M. H. Elmirghani[2]
[1]School of Electronic and Electrical Engineering, University of Leeds, LS2 9JT, United Kingdom
[2]Department of Engineering, Faculty of Natural, Mathematical & Engineering Sciences Kings College London
elkal@leeds.ac.uk, {ahmad.qidan, taisir.elgorashi, jaafar.elmirghani}@kcl.ac.uk

*Abstract*—Optical wireless communication (OWC) provides high aggregate data rates in the range of Terabits per second (Tb/s). Specifically, OWC using infrared lasers as transmitters has been considered as a strong candidate in the next generation of wireless communication. Rate splitting (RS) is a transmission scheme derived to improve spectral efficiency in dense wireless networks. In RS, the transmitted power is allocated to different messages, common and private messages, serving multiple users simultaneously, where each user can decode the desired message following a certain procedure. Moreover, two-tier precoding RS scheme has been proposed to overcome the limitations of traditional RS in multi-group scenarios. In this context, power allocation (PA) is a crucial issue, which can affect the performance of RS. Therefore, we formulate a PA optimization problem to enhance the data rates of RS-based OWC networks. However, such optimization problems are complex due to the use of different messages intended to the users. In this paper, we design and train a deep neural network (DNN) model to determine the power allocated to the messages of RS, while fulfilling the demands of users. The results show the accuracy of our trained DNN model when used in an online phase.

## I. Introduction

Optical wireless communication (OWC) has been considered as a strong candidate in the next generation of wireless communications due to its massive licence-free bandwidth, high secrecy, and high spectral and energy efficiency [1], [2] . Studies for more than a decade on light-emitting diodes (LEDs) have shown their capabilities in transmitting information at gigabit-per-second (Gbps) communication speeds. Despite the features of LEDs as transmitters, they suffers from low modulation speeds. Additionally, LEDs are used primarily for providing illumination. Therefore, the deployment of a high number of LEDs to offer wide coverage area and seamless user-transition may contradict with the recommended illumination levels. In this context, infrared lasers, for example vertical-cavity surface-emitting (VCSEL) lasers, were used in [3], [4], [5] to serve users at Terabit-per-second (Tbps) communication speeds. VCSELs are characterised by their high modulation speed, spectral purity, and low cost. However, the total transmitted power of the VCSEL must adhere to the eye safety constraints [6].

Multi-user Interference in optical wireless networks can be managed using transmission techniques designed for radio frequency (RF) wireless networks [7] [8]. However, those techniques must be modified considering the unique characteristics of the optical signal [3], [9]. In [3], rate splitting (RS) was applied in multiple-input single-output (MISO) laser-based OWC scenarios to serve multiple users, and therefore, enhancing the spectral efficiency. Basically, RS aims to relax the requirements of channel state information (CSI) at the transmitters compared to other transmit precoding schemes. The methodology of RS relies on splitting the message of the user into common and private messages while allocating different power levels to these messages.

Despite the high spectral efficiency of RS [3], its performance suffers considerable degradation in large scale OWC networks due to noise enhancement. Given this, a two-tier precoding RS scheme referred to as hierarchical rate splitting (HRS) was proposed in [10] to tackle the challenges of RS in OWC networks. In HRS, the optical access points (AP) transmit three different messages, i.e., inner, outer and private messages, to manage inter-group interference, after arranging the users into multiple groups, and intra-group interference as joint problems. In [10], our previous work, a power allocation optimization problem was formulated to optimize the performance of HRS in a laser based OWC. The optimization problem is defined as a non-convex problem, which is not easy to solve, due to the nature of HRS. Therefore, the successive convex approximation algorithm was proposed to provide sub-optimal solutions. However, such algorithm might fail to provide instantaneous solutions to meet high quality of service (QoS) requirements, especially in highly dynamic OWC scenarios where solutions might become outdated. Therefore, the use of artificial intelligence (AI)- based techniques to solve such complex optimization problems is essential.

Recently, deep learning (DL) models have been considered to provide practical solutions for NP-hard optimization problems. DL models must be trained to interact with the environment to optimize a certain metric. In the context of power allocation, several studies have adopted DL algorithms to enhance the performance of RF wireless networks [11], [12]. In [11], the authors proposed a fully connected deep neural network (DNN) to approximate the weighted minimum mean square error (WMMSE), as it is a popular power allocation algorithm, and the simulation results demonstrate that the proposed DNN can achieve fairly accurate approximation with low computational time . For achieving higher performance and magnitude speedup in computational time in RF networks [12], a parallel DNN architecture was proposed to optimize licensed power allocation and unlicensed time fraction. Moreover, a real-time sub-optimal solution has been achieved by using AI-based algorithms for optimizing resource allocation and maximizing the overall sum rate considering blind multiple access scheme in laser-based optical wireless networks [13]. However, the optimization of power allocation using AI-based techniques in optical and RF wireless networks is still under investigation.

In this paper, we design and train a DNN model to optimize power allocation in a real time RS-based OWC network. Our network is composed of multiple optical APs deployed to serve a number of users with different demands. We first define the system model and derive the user rate considering the application of RS. Then an objective function is derived to maximize the sum rate of the network under several constraints that guarantee consuming power lower or equal to the available power budget, while users experiencing high QoS. Finally, a DNN model is designed and trained to estimate a sub-optimal power allocation among the messages generated as a result of using RS. The results show the practicality and accuracy of our trained DNN model in solving complex optimization problem with solutions significantly close to the optimal ones.

## II. SYSTEM MODEL

Fig. 1 shows the system model of an RS laser-based optical wireless network. The users denoted by $k, k = \{1,..,K\}$, are randomly distributed on the receiving plane, and are served by $L$, optical Aps deployed on the ceiling to provide uniform coverage. Each optical AP is denoted by $l$, $l = \{1,..,L\}$ and is composed of multiple VCSELs. Each user is equipped with an optical receiver, which has $M$ photodiodes, i.e. angle diversity receiver (ADR) that point to distinct directions in order to provide linearly independent channel responses. The direction of each photodiode is specified by the azimuth and elevation angles and has a narrow Field of View (FoV) [14]. Thus, a connection to most of the available APs can be ensured, i.e., full connectivity. Given that, the transmitted signal can be written in a vector form as

$$\mathbf{x} = [x_1 \quad x_2 \quad ... \quad x_L]^T \in \mathbb{R}_+^{L \times 1}, \quad (1)$$

where $x_l$ denotes the signal transmitted by optical AP $l$. Without considering any multiple access schemes, the signal received by user k can be written as:

$$y^{[k]} = \mathbf{h}^{[k]}(m^{[k]})^T \mathbf{x} + z^{[k]} \quad (2)$$

where $\mathbf{h}^{[k]}(m^{[k]})^T$ is the channel vector of user $k$ at photodiode $m$, $m \in M$, $z^{[k]}$ is defined as real valued additive white Gaussian noise with zero mean and variance given by the sum of shot noise, thermal noise and the intensity noise of the laser. In this work, all the optical APs are connected through a central unit (CU). The CU has information to optimize resource allocation, and CSI is available at the transmitters to design the precoding matrices of the users according to the methodology of RS to guarantee multi-user interference management.

### A. VCSEL Transmitter

We consider using the VCSEL as a transmitter due to its high modulation speed, low cost and high energy efficiency [6]. Generally, the VCSEL has Gaussian beam profile, and its power distribution can be determined by three factors, the wavelength $\lambda$, the beam waist $W_0$ and the distance $d$ between the transmitter and user. In this context, the beam radius of the VCSEL at photodiode $m$ of user $k$ can be calculated as:

$$W_d = W_0 \left(1 + \left(\frac{d}{d_{Ra}}\right)^2\right)^{\frac{1}{2}} \quad (3)$$

where $d_{Ra}$ is the Rayleigh distance. Furthermore, the spatial intensity profile of the VCSEL transmitter over the transverse plane at distance $d$ is given by

$$I(r,d) = \frac{2P_t}{\pi W_d^2} \exp\left(-\frac{2r^2}{W_d^2}\right) \quad (4)$$

where $P_t$ is the optical power and $r$ is the radial position. Assuming that a user is located right under the VCSEL, the received power at its photodiode $m$ can be calculated as

$$P_m = \int_0^{r_m} I(r,d) 2\pi r dr = P_t \left[1 - \exp\left(\frac{-2r_m^2}{W_d^2}\right)\right] \quad (5)$$

where $r_m$ is the radius of photodiode $m$. The detection area of photodiode $m$ determined by $A_m = \frac{A_{rec}}{M}$, $A_{rec}$ is the whole detection area of the receiver, $m \in M$. More scenarios and mathematical expressions of the received power of the laser can be found in details in [3].

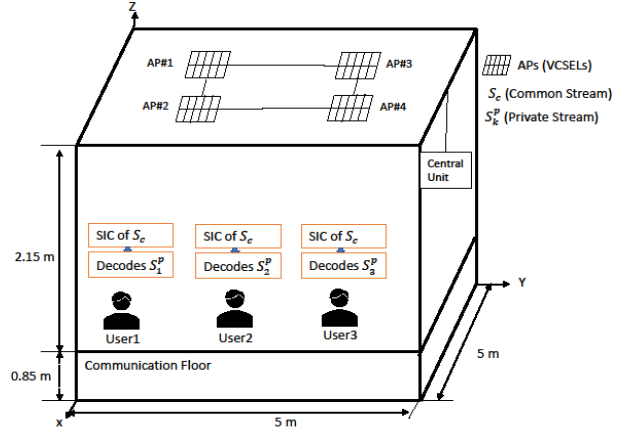

Fig 1. An OWC system model, and a RS illustrative example.

### B. Rate Splitting

RS is a transmission scheme that can be applied in multiple input-single output (MISO) broadcast channel (BC) networks to manage multi-user interference with the need for channel state information (CSI) at the transmitters. The basic idea of RS is to split the message of each user $k$ into common and private messages, generating a super common message superimposed on top of the private messages. The super common message is derived from a public codebook, and all users can decode it with a minimum error. RS can serve $K$ users simultaneously, and each user after decoding the common message can decode its useful information following a specific interference cancelation methodology. However, the performance of RS suffers performance degradation in dense wireless network, more specifically OWC networks, due to severe noise resulting from interference cancelation [2].

One of the solutions for enhancing the performance of RS in OWC is to form $G$ groups of users, where each group contains a set of users $K_g$, $\sum_{g=1}^{G} K_g = K$. Note that, partitioning users into multiple groups can be based on distance using the well-known K-means clustering algorithm. In [10], HRS, i.e., two-tier of RS, is proposed to manage the interference among groups, i.e., inter-group interference, and the interference among the users belonging to each group, i.e., intra-group interference. Therefore, two precoding matrices are used, which are denoted by the outer precoder $\mathbf{B}_g$ and inner precoder $\mathbf{W}_g = [\mathbf{w}_{g1}, \mathbf{w}_{g2} \dots \dots, \mathbf{w}_{gK}]$. In this context, all the transmitters send an outer-common message $s_{oc}$ superimposed over inner-common messages $\sum_{g=1}^{G} s_{ic,g}$ and private messages $\sum_{k=1}^{K_g} s_{gk}$ intended to the groups formed in the network. Therefore, the transmitted signal of HRS can be expressed as

$$\mathbf{x} = \sqrt{P_{oc}} \mathbf{w}_{oc} s_{oc} + \sum_{g=1}^{G} \mathbf{B}_g \left( \sqrt{P_{ic,g}} \mathbf{w}_{ic,g} s_{ic,g} + \sum_{k=1}^{K_g} \sqrt{P_{gk}} \mathbf{w}_{gk} s_{gk} \right) \quad (6)$$

where $P_{oc}$ is the power of the outer common message that can be determined by $P_{oc} = P(1 - \beta)$, where $\beta \in (0,1]$ is the fraction of the total power allocated to the messages of each group. Moreover, $P_{ic,g}$ is the power of the inner common message intended to group $g$ determined by $P_{ic,g} = \frac{P\beta}{G}(1 - \alpha)$, where $\alpha \in (0,1]$ determines the fraction of power allocated to the private messages of each group. The last notation of power is $P_{gk}$ for the private message intended to user $k$ in group $g$, where $P_{gk} = \frac{P\beta}{K}\alpha$. Therefore, the received signal of a generic user $k$ in group $g$ is given by

$$y_{gk} = \sqrt{P_{oc}}\, \mathbf{h}_{gk}^H \mathbf{w}_{oc} s_{oc} + \sqrt{P_{ic,g}}\, \mathbf{h}_{gk}^H \mathbf{B}_g \mathbf{w}_{ic,g} s_{ic,g}$$
$$+ \sqrt{P_{gk}} \mathbf{h}_{gk}^H \mathbf{B}_g \mathbf{w}_{gk} s_{gk} + \sum_{j \neq k}^{K_g} \sqrt{P_{gj}} \mathbf{h}_{gk}^H \mathbf{B}_g \mathbf{w}_{gj} s_{gj} \quad (7)$$
$$+ \sum_{l \neq g}^{G} \mathbf{h}_{gk}^H \mathbf{B}_l \mathbf{W}_l \mathbf{P}_l \mathbf{s}_l + z_{gk}$$

where $\mathbf{h}_{gk}^H$ is the channel response vector between user $k$ within group $g$ and the $L$ optical transmitters, $z_k$ is defined in equation (2). The decoding procedure of user $k$ in hierarchical RS follows three steps. Firstly, user $k$ decodes the outer common message by treating all other messages, inner common messages and private messages as noise. Secondly, the inner common message is decoded while treating the remaining messages as noise. Note that, each user is able to perform successive interference cancellation (SIC) to remove the decoded signals. Finally, user $k$ decodes its private message by treating all other private messages as noise. From the previous three steps, the Signal to Interference plus Noise Ratios (SINRs) for all the messages can be expressed as

$$\gamma_{gk}^{oc} = \frac{P_{oc}\left|\mathbf{h}_{gk}^H \mathbf{w}_{oc}\right|^2}{\sum_{l=1}^{G}\sum_{j=1}^{K_g} P_p^{[j,l]}\left|\mathbf{h}_{gk}^H \mathbf{B}_l \mathbf{w}_{lj}\right|^2 + \sum_{l=1}^{G} P_{ic}^{[l]}\left|\mathbf{h}_{gk}^H \mathbf{B}_l \mathbf{w}_{ic,l}\right|^2 + \sigma_{gk}^2} \quad (8)$$

$$\gamma_{gk}^{ic} = \frac{P_{ic}^{[g]}\left|\mathbf{h}_{gk}^H \mathbf{B}_g \mathbf{w}_{ic,g}\right|^2}{\sum_{l=1}^{G}\sum_{j=1}^{K_g} P_p^{[j,l]}\left|\mathbf{h}_{gk}^H \mathbf{B}_l \mathbf{w}_{lj}\right|^2 + \sum_{l \neq g}^{G} P_{ic}^{[l]}\left|\mathbf{h}_{gk}^H \mathbf{B}_l \mathbf{w}_{ic,l}\right|^2 + \sigma_{gk}^2} \quad (9)$$

$$\gamma_{gk}^{p} = \frac{P_p^{[k,g]}\left|\mathbf{h}_{gk}^H \mathbf{B}_g \mathbf{w}_{gk}\right|^2}{\sum_{l=1}^{G}\sum_{j=1}^{K_g} P_p^{[j,l]}\left|\mathbf{h}_{gk}^H \mathbf{B}_l \mathbf{w}_{lj}\right|^2 - \Lambda + \sum_{l \neq g}^{G} P_{ic}^{[l]}\left|\mathbf{h}_{gk}^H \mathbf{B}_l \mathbf{w}_{ic,l}\right|^2 + \sigma_{gk}^2} \quad (10)$$

where $\gamma_{gk}^{oc}$ is the SINR of the outer common message, $\gamma_{gk}^{ic}$ is the SINR of the inner common message and $\gamma_{gk}^{p}$ is the SINR of the private message. Moreover, $\Lambda$ in equation (10) is given by $P_p^{[k,g]}\left|\mathbf{h}_{gk}^H \mathbf{B}_g \mathbf{w}_{gk}\right|^2$. Accordingly, the sum rate of the hierarchical RS can be calculated as

$$R_{\text{sum}}^{HRS} = R_{oc}^{HRS} + R_{ic}^{HRS} + R_p^{HRS}, \quad (11)$$

where $R_{oc}^{HRS}$ is the achievable rate of the outer common message given by

$$R_{oc}^{HRS} = \log_2(1 + \gamma^{oc}),$$
$$\gamma^{oc} = \min_{g,k}\{\gamma_{gk}^{oc}\}, \quad (12)$$

$R_{ic}^{HRS}$ represents the sum rate of the inner common messages is given as

$$R_{ic}^{HRS} = \sum_{g=1}^{G} \log_2(1 + \gamma_g^{ic}),$$
$$\gamma_g^{ic} = \min_{k}\{\gamma_k^{ic}\} \quad (13)$$

Note that, $\gamma^{oc} = \min_{g,k}\{\gamma_{gk}^{oc}\}$ and $\gamma_g^{ic} = \min_{k}\{\gamma_k^{ic}\}$ are conditions to ensure the decodability of the outer and inner common messages with minimum error. Finally, $R_p^{HRS}$ is the sum rate of the private messages given by

$$R_p^{HRS} = \sum_{g=1}^{G} \sum_{k=1}^{K_g} \log_2(1 + \gamma_{gk}^{p}) \quad (14)$$

### III. OPTIMIZATION PROBLEM

The application of HRS in OWC networks results in complexity in terms of power allocation due to the use of three different messages. In other words, the powers of the outer common, inner common and private messages are coupled (see equations (8)-(10)). Given that, a method for managing the power allocated to each message is needed in such interference management schemes. In this section, a power allocation optimization problem is formulated considering the methodology of two-tier RS such that the sum rate of the network is maximized, and the user experience of service is enhanced. As previously mentioned, the users in the network are divided into multiple groups, and each group contains a unique set of users that are close to each other, i.e., K-means clustering. It is worth mentioning that the outer common message in HRS is generated to manage inter-group interference. In high SINR scenarios, the interference among the groups can be assumed to be constant, and therefore, a fixed power can be allocated to the outer common message to ease the implementation of HRS. In this context, the powers of the inner common and private message are defined as vital parameters that can highly dictate the performance of the network. Therefore, we define a utility-based objective function that aims to maximize the sum rate of the users taking into the consideration the power allocated to these messages. At this point, the objective function of the sum rate is expressed as follows:

$$U\left(R_{sum}\left(P_{oc}^{\dagger}, P_{ic}, P_{p}\right)\right)$$
$$= R_{oc}\left(P_{oc}^{\dagger}, P_{ic}, P_{p}\right)$$
$$+ \sum_{g}^{G}\left(R_{ic}^{[g]}(P_{ic}, P_{p})\right.$$
$$\left.+ \sum_{k}^{K_g} R_{p}^{[k,g]}(P_{ic}, P_{p})\right), \quad (15)$$

where $P_{oc}^{\dagger}$ is a fixed power allocated to the outer common message, and $R_{oc}(P_{oc}^{\dagger}, P_{ic}, P_p)$, $R_{ic}^{[g]}(P_{ic}, P_p)$ and $R_p^{[k,g]}(P_{ic}, P_p)$ can be easily derived from equations (8)-(14). Note that, $U(\cdot)$ is defined as a logarithmic function, i.e. $U(\cdot) = \log(.)$ to achieve a proportional fairness among the users belonging to different groups using the HRS scheme. The optimization problem for maximizing the sum rate of the $K$ users in the network can be formulated under the power and high QoS constraints as follows:

$$\max_{p} U\left(R_{sum}(P_{oc}^{\dagger}, P_{ic}, P_p)\right) \quad (16)$$

s.t. $R_{sum}(P_{oc}^{\dagger}, P_{ic}, P_p) \geq R_{min}(P_{oc}^{\dagger}, P_{ic}, P_p)$ (16a)

$$P_{oc}^{\dagger} + \sum_{g \in G}\left(P_{ic}^{[g]} + \sum_{k \in K_g} P_p^{[k,g]}\right) \leq P_T \quad (16b)$$
$$, \forall g \in G$$

$$P_{ic}^{[g]} + \sum_{k \in K_g} P_p^{[k,g]} \leq P_{max}^{[g]}, \forall g \in G \quad (16c)$$

$$P_p^{[k,g]} \leq P_p^T, \forall k \in K_g \quad (16d)$$

$$P_{oc}^{\dagger}, P_{ic}^{[g]}, P_p^{[k,g]} \geq 0, \forall g \in G, \forall k \in K_g \quad (16e)$$

The first physical constraint guarantees that the sum rate of the network is higher than the minimum data rate required to ensure high QoS under the use of two-tier RS. The second constraint ensures that the total power consumed by the network is less than or equal to the maximum power consumption $P_T$ allowed in the network, which is determined according to eye safety regulations [6]. The third constraint controls the total power allocated to each group $P_{max}^{[g]}$. The last constraint limits the power allocated to the private message intended to each user, which must be less than or equal to $P_p^T$, which denotes the maximum power that can be consumed by each user. The problem in (16) is defined as a non-convex problem, which is difficult to solve, and it has high computational complexity that increases with the size of the network. In [5], our previous work, a parametric approach is developed to solve optimization problems of similar nature at a given parameter, providing sub-optimal solutions at a relatively low complexity compared to some deterministic algorithms.

However, the solutions of such algorithms are expected to be outdated in real-time scenarios due to the fact that the circumstances of the wireless networks tend to change very frequently. Given that, ML models are highly needed to provide instantaneous solutions that can meet the highly varying user demands. the.

In the following, we design an DNN model, and train it to learn the optimal power allocation decisions for RS-based OWC networks in order to provide fast and effective solutions close to the optimal ones.

## IV. DEEP NEURAL NETWORK

DNN is a ML technique that can be trained to solve various optimization problems. An DNN model consists of an input layer, multiple hidden layers and an output layer, where each layer has multiple neurons. Interestingly, the setup of the DNN model is mainly based on two functions referred as an activation function and a linear transformation function. Firstly, the activation function works on each neuron to calculate its output, and a neuron can be deactivated if its output is irrelevant. Secondly, the linear transformation function is defined based on the model of the DNN considered to solve an optimization problem, which is the convolution function in this work [13]. Our aim in (16) is to find the optimal power levels allocated to the inner common and private messages in order to maximize the sum rate of the two-tier RS in OWC. In this context, the DNN must be trained to find an optimal power allocation vector in real time under the constraints of power budget and requirements of high QoS. The implementation of the DNN model involves two steps as follows.

### A. DNN offline phase:

In this step, a dataset must be generated from solving the main optimization problem in (16). Note that, generating a large size dataset involves also high complexity. However, it is an offline process. After that, the training must occur over a set of iterations to find the unknown mapping $f(\mathcal{V};.)$ between the input layer and output layer of the DNN model. In our DNN, users send their QoS requirements through a WiFi AP connected to the CU of the network. The training process helps determine a set of optimal weight $\mathcal{V}^*$ that can provide sub-optimal solutions once the application of the DNN model is considered in an online phase to estimate $\widehat{\mathbf{P}} = \left[\hat{P}^{[1]}, \dots, \hat{P}^{[g]}, \dots, \hat{P}^{[G]}\right]$, where $\hat{P}^{[g]} = \hat{P}_{ic}^{[g]} + \sum_{k \in K_g} \hat{P}_p^{[k,g]}$ is the estimation of the power allocated to the users belonging to each group $g$. Note that, the output layer of our DNN has a size of $K + G + 1$ due to the fact that our DNN must learn the powers allocated to the inner common and private messages in addition to the power budget. Moreover, $\widehat{\mathbf{P}}$ is estimated according to the optimal power vector $\mathbf{P}^* = \left[P^{*[1]}, \dots, P^{*[g]}, \dots, P^{*[G]}\right]$, where $P^{*[g]} = P_{ic}^{*[g]} + \sum_{k \in K_g} P_p^{*[k,g]}$.

Let us assume that our training dataset contains $N$ data points, where each data point $n$ corresponds to $K$ users sending their QoS requirements. For the training input denoted as v(n), the optimal power allocation vector is given by $\mathbf{P}^*(n)$ while the estimation of power for this data point is given by $\widehat{\mathbf{P}}(n)$. At this point, the training process must occur over a number of iterations to minimize a certain loss function between the optimal and estimated power allocation as follows:

$$\min_{\mathbf{W}} \frac{1}{N} \sum_{n=1}^{N} \ell\left(\widehat{\mathbf{P}}(n), \mathbf{P}^*(n)\right), \qquad (17)$$

where $\mathbf{W}$ is the weight, and $\ell(.,.)$ is the root mean square error (RMSE) function. After completing the training stage, our trained DNN must not be overfitted, and can be used to find sub-optimal solutions in scenarios that are not even included in the training dataset.

*B. DNN online application:*

After generating the dataset and training the DNN model in an offline phase, its use is considered at the optical APs to perform instantaneous data rate-maximization by finding the sub-optimum power allocation. Basically, users send their data rate requirements to the optical APs through uplink transmission. Subsequently, this information is injected into the trained DNN model to perform power allocation among the users considering the formation of multiple groups and methodology of two-tier RS. Note that, the DNN model estimates a new power allocation vector every time the users send a new set of demands.

## V. SIMULATION RESULTS

We consider an indoor environment with the dimensions 5 m × 5 m × 3 m to evaluate the performance of the trained DNN model compared with some benchmark schemes. On the ceiling, $L = 4$ optical APs are deployed, and each AP is composed of multiple transmitters to expand the coverage. On the receiving plane, $K=6$ users are distributed randomly, and arranged into multiple groups. Recall that, each user is equipped with ADR that provides a wide FoV, ensuring full connectivity to most of the available optical APs. The rest of the parameters are mentioned in Table I.

The dataset consists of 10000 samples, generated by the simulation of RS multi-user scenarios in OWC system. Three datasets with different sizes are obtained from the main dataset. The training dataset contains 60% of the data points, the validation dataset contains 20%, and the last 20% is used for the testing dataset. The validation dataset is not used to update the network weights, however, used for unbiased evaluation that determined how well our DNN learns while training.

In Fig. 2, the sum rate of the network is shown against different values for the laser beam waist $W_0$. It can be seen that the sum rate of the network increases with the beam waist regardless of the scheme considered, due to the fact that more power is received by the users. The figure also shows that the performance of the optimization problem in (16) results in enhancing the sum rate of the users considerably compared to the performance of the conventional HRS in which the power is simply allocated uniformly among the messages intended to the users. Note that, our proposed DNN-model shows high accuracy in estimating power allocation in all the scenarios considered, and achieves significant solutions close to the optimal ones obtained by solving the optimization problem in (16).

In Fig. 3, the sum rate of the proposed model is depicted against the number of users. It is shown that our proposed model achieves higher sum rates compared to the conventional RS and HRS schemes, and other traditional scheme referred to as Orthogonal Multiple-Access (OMA). Note that, the sum rates of all the schemes increase with the number of users due to the fact that the aggregate data rates are determined, and the users experience different channel gains. However, the lack of resources in OMA and the high noise caused by interference cancellation due to the transmission to all users in conventional RS must be taken into consideration. It is worth mentioning that the proposed model divides the users into multiple groups, and then, uses a certain methodology for interference management.

Table I. Simulation Parameters

| Laser-based OWC parameter | Value |
|---|---|
| Laser Beam waist, $W_0$ | $5 - 20 \ \mu m$ |
| Laser Bandwidth | 5 GHz |
| Laser Wavelength | 850 $nm$ |
| Receiver Responsivity | 0.4 A/W |
| Receiver FOV | 45 deg |
| Area of the photodetector | 15 mm$^2$ |
| Receiver noise current spectral density | 4.47 pA/√Hz |
| **DNN parameter** | **Value** |
| Model | convolutional |
| Number of hidden layers | 4 |
| Dataset size | 10000 |
| Training | 60% of dataset |
| Validation | 20% of dataset |

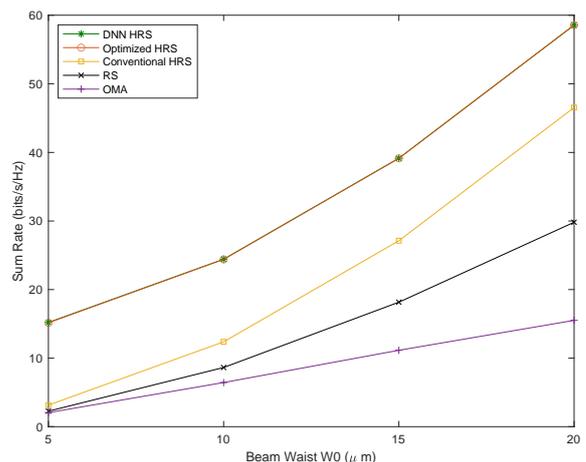

Fig 2. Sum rates versus the beam waist.

In Fig. 4, the sum rate of the network is shown against a range of SNR values from 5 dB to 35 dB using the proposed model, which allocates the power among messages of the two-tier RS with the aim of maximizing the sum rate of the network. Note that, the sum rates of all the schemes increase with the SNR, where each user can successfully decode its desired information at high SNR values. The figure further shows that allocating the power in an optimal fashion result in higher sum rates compared to the other schemes, conventional RS, HRS and OMA. Moreover, the proposed model provides relatively high sum rates in low SNR scenarios, which can be caused by many reasons including the imperfection of CSI at the transmitters.

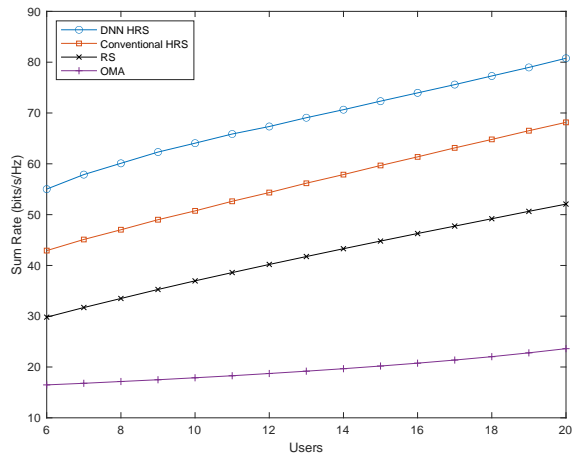

Fig 3. Sum rates versus $K$ users.

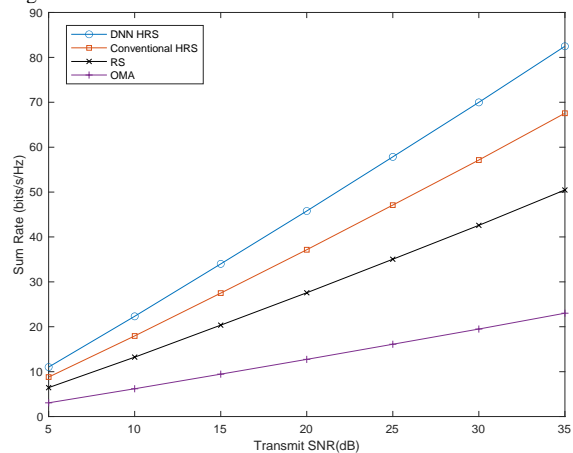

Fig 4. Sum rates versus SNR.

## VI. CONCLUSIONS

In this paper, we design a DNN model to solve the power allocation optimization problem in a two-tier RS OWC network. We first define the system model to be composed of multiple lasers-based APs serving a number of users. Then, the two-tier RS is defined in details, and the sum rate of the network is derived. Finally, an optimization problem is formulated to maximize the sum rate of the network under the constraints of power budget and high QoS. To provide practical and instantaneous solutions, a DNN model is designed and trained over a dataset generated by solving the power allocation problem. The results show that the proposed model has high accuracy in providing significant solutions close to the optimal ones, and it has the ability to cope with high density and low SNR OWC scenarios compared to the conventional RS, HRS and OMA.

## ACKNOWLEDGMENTS

The authors would like to acknowledge funding from the Engineering and Physical Sciences Research Council (EPSRC) INTERNET (EP/H040536/1), STAR (EP/K016873/1) and TOWS (EP/S016570/1) projects. Khulood would like to thank King Abdulaziz University in the Kingdom of Saudi Arabia for funding her PhD scholarship. All data are provided in full in the results section of this paper.

## REFERENCES

[1] H. Elgala, R. Mesleh, and H. Haas, "Indoor optical wireless communication: Potential and state-of-the-art," IEEE Communications Magazine, vol. 49, no. 9, pp. 56–62, 2011, doi: 10.1109/MCOM.2011.6011734.

[2] A. Adnan-Qidan, M. Morales-Cespedes, and A. G. Armada, "User-centric blind interference alignment design for visible light communications," IEEE Access, vol. 7, pp. 21 220–21 234, 2019.

[3] K. Alazwary, A. A. Qidan, T. El-Gorashi, and J. M. H. Elmirghani, "Rate splitting in VCSEL-based optical wireless networks," 2021 6th International Conference on Smart and Sustainable Technologies, SpliTech 2021, pp. 1–5, 2021, doi: 10.23919/SpliTech52315.2021.9566354.

[4] A. Qidan, T. El-Gorashi, and J. Elmirghani, "Towards Terabit LiFi Networking," no. Photoptics, pp. 203–212, 2022, doi: 10.5220/0010955000003121.

[5] A. A. Qidan et al., "Multi-User Rate Splitting in Optical Wireless Networks," pp. 1–37, 2022, [Online]. Available: http://arxiv.org/abs/2207.11458

[6] M. D. Soltani et al., "Safety Analysis for Laser-based Optical Wireless Communications: A Tutorial," pp. 1–54, 2021, [Online]. Available: http://arxiv.org/abs/2102.08707

[7] G. Zhou, Y. Mao, and B. Clerckx, "Rate-Splitting Multiple Access for Multi-Antenna Downlink Communication Systems: Spectral and Energy Efficiency Tradeoff," IEEE Trans Wirel Commun, vol. 21, no. 7, pp. 4816–4828, 2022, doi: 10.1109/TWC.2021.3133433.

[8] Y. Mao, B. Clerckx, and V. O. K. Li, "Energy Efficiency of Rate-Splitting Multiple Access, and Performance Benefits over SDMA and NOMA," Proceedings of the International Symposium on Wireless Communication Systems, vol. 2018-Augus, no. Ic, pp. 1–5, 2018, doi: 10.1109/ISWCS.2018.8491100.

[9] S. Naser et al., "Rate-Splitting Multiple Access: Unifying NOMA and SDMA in MISO VLC Channels," IEEE Open Journal of Vehicular Technology, vol. 1, no. December, pp. 393–413, 2020, doi: 10.1109/OJVT.2020.3031656.

[10] K. Alazwary, A. A. Qidan, T. El-Gorashi, and J. M. H. Elmirghani, "Optimizing Rate Splitting in Laser-based Optical Wireless Networks," 2022 IEEE International Conference on Communications Workshops, ICC Workshops 2022, pp. 463–468, 2022, doi: 10.1109/ICCWorkshops53468.2022.9814503.

[11] H. Sun, X. Chen, Q. Shi, M. Hong, X. Fu, and N. D. Sidiropoulos, "Learning to Optimize: Training Deep Neural Networks for Interference Management," IEEE Transactions on Signal Processing, vol. 66, no. 20, pp. 5438–5453, 2018, doi: 10.1109/TSP.2018.2866382.

[12] G. Chai, W. Wu, Q. Yang, R. Liu, and K. S. Kwak, "Learning to optimize for resource allocation in LTE-U networks," China Communications, vol. 18, no. 3, pp. 142–154, 2021, doi: 10.23919/JCC.2021.03.012.

[13] A. A. Qidan, T. El-Gorashi, and J. M. H. Elmirghani, "Artificial Neural Network for Resource Allocation in Laser-based Optical wireless Networks," IEEE International Conference on Communications, vol. 2022-May, pp. 3009–3015, 2022, doi: 10.1109/ICC45855.2022.9839259.

[14] A. A. Qidan, M. Morales-Cespedes, T. El-Gorashi, and J. M. H. Elmirghani, "Resource Allocation in Laser-based Optical Wireless Cellular Networks," 2021 IEEE Global Communications Conference, GLOBECOM 2021 - Proceedings, 2021, doi: 10.1109/GLOBECOM46510.2021.9685357.

[15] B. R. Marks and G. P. Wright, "Technical Note—A General Inner Approximation Algorithm for Nonconvex Mathematical Programs," https://doi.org/10.1287/opre.26.4.681, vol. 26, no. 4, pp. 681–683, Aug. 1978, doi: 10.1287/OPRE.26.4.681.